\documentclass[twocolumn,showpacs,superscriptaddress,prb,floatfix]{revtex4}
\usepackage{graphicx}
\usepackage{epsfig}
\usepackage[usenames]{color}
\usepackage{subeqn}
\usepackage{ulem} 

\newcommand{\ket}[1]{\ensuremath{|#1\rangle}}
\newcommand{\bracket}[2]{\ensuremath{\langle #1|#2\rangle}}

\newcommand{\bk}{{\bf k}}
\newcommand{\br}{{\bf r}}
\newcommand{\bt}{{\bf t}}
\newcommand{\bu}{{\bf u}}

\newcommand{\bS}{{\bf S}}
\newcommand{\bT}{{\bf T}}

\date{\today}

\begin{document}
\title{Interaction of Lamb modes with two-level systems in amorphous
  nanoscopic membranes} 
\author{T. K{\"u}hn}
\affiliation{Nanoscience Center, Department of Physics, University of
  Jyv\"askyl\"a, P.O. Box 35, FIN-40014 University of Jyv\"askyl\"a, Finland} 

\author{D. V. Anghel}
\affiliation{Department of Theoretical Physics, National Institute for
  Physics and Nuclear Engineering--''Horia Hulubei'', Str. Atomistilor
  no.407, P.O.BOX MG-6, Bucharest - Magurele, Romania} 
\affiliation{Bogoliubov Laboratory of Theoretical Physics, JINR Dubna, Russia} 
\author{Y. M. Galperin}
\affiliation{Department of
Physics \& Centre of Advanced Materials and Nanotechnology, University
of Oslo, PO Box 1048 Blindern, 0316 Oslo,~Norway}
\affiliation{Argonne National Laboratory, 9700 S. Cass
Av., Argonne, IL 60439, USA}
\affiliation{A. F. Ioffe
Physico-Technical Institute of Russian Academy of Sciences, 194021
St. Petersburg, Russia}
\author{M. Manninen}
\affiliation{Nanoscience Center, Department of Physics, University of Jyv\"askyl\"a, P.O. Box 35, FIN-40014 University of Jyv\"askyl\"a, Finland}
\begin{abstract}\noindent
Using a generalized model of interaction between a two-level system (TLS) and 
an arbitrary deformation of the material, we calculate the interaction of 
Lamb modes with TLSs in amorphous nanoscopic membranes. We compare the 
mean free paths of the Lamb modes with different symmetries and 
calculate the heat conductivity $\kappa$. In the limit 
of an infinitely wide membrane, the heat conductivity is divergent. 
Nevertheless, the finite size of the membrane imposes a lower cut-off 
for the phonons frequencies, which leads to the temperature dependence  
$\kappa\propto T(a+b\ln T)$. This temperature dependence is a
hallmark of the TLS-limited heat conductance at low temperature.
\end{abstract}
\maketitle

\section{Introduction} \label{sec_introduction}

The development of nanodetectors and the strict requirements on their 
performance triggered intense experimental and theoretical studies of 
their thermal properties. These detectors work usually in  
a temperature range around 1~K or below and are 
supported by thin, insulating membranes. 
The thickness of such membranes is of the order of 100 nm, which, in 
the given temperature range, makes it comparable to the dominant thermal 
phonon wavelength. In problems where the phonon wavelength is comparable 
to or longer than some of the dimensions of the system in question, the three 
dimensional (3D) phonon gas model cannot be applied anymore to calculate 
the system's thermal properties. Instead, one 
has to use the phonon modes specific to the system, which 
are the eigenmodes of the elastic equation for the given geometry.

The membranes that support the detectors are made of amorphous, low stress 
silicon-nitride (SiN$_\mathrm{x}$) and their thermal properties have been measured 
in various geometries by different groups (see for example 
Refs.\ [\onlinecite{Leivo:thesis,ApplPhysLett.72.1305.Leivo,
ApplPhysLett.72.2250.Holmes,PhysicaB.284.1968.Woodcraft,ApplPhysLett86.251903.Hoevers,
SolidStateComm.129.199.ZinkHellman}]).  
Depending on the quality and the dimensions of the samples, and possibly 
the temperature range in which measurements were done, the heat flux 
along the membrane may be due to either diffusive \cite{ApplPhysLett.72.1305.Leivo,
PhysRevLett.81.2958.Anghel,ApplPhysLett.72.2250.Holmes,PhysicaB.284.1968.Woodcraft} 
or radiative phonon transport.\cite{ApplPhysLett.72.2250.Holmes,ApplPhysLett86.251903.Hoevers} 
In the case of diffusive phonon transport, it was observed that the 
heat conductivity $\kappa$ is roughly proportional to $T^2$. 

\begin{figure}
\begin{center}
\resizebox{70mm}{!}{\includegraphics{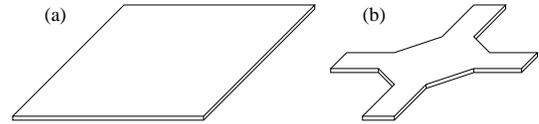}}
\caption{(a) Full, dielectric membrane. (b) Cut membrane 
for better thermal insulation.}\label{membrane2}
\end{center}
\end{figure}

For a better thermal insulation of the detector, the underlying membrane 
is sometimes cut. The result is a self-supporting structure with a 
wider area in the middle, connected to the bulk by long, narrow 
bridges, like in Fig. \ref{membrane2}b.\cite{Leivo:thesis,ApplPhysLett.72.1305.Leivo}
The heat conductivity along such bridges has again a power law dependence on 
the temperature, $\kappa\propto T^p$, where $p$ takes values between 
1.5 and 2.\cite{Leivo:thesis,ApplPhysLett.72.1305.Leivo} For the samples 
measured in [\onlinecite{ApplPhysLett.72.1305.Leivo}], $p$ increased with 
the width of the bridge.

The heat capacity of a membrane is more difficult to measure 
directly,  since the membrane is always in contact with the bulk. 
However, it can be estimated by applying AC heating and measuring
the amplitude of the temperature 
oscillations. In this way Leivo and Pekola \cite{ApplPhysLett.72.1305.Leivo} 
observed that the ratio $c_V/\kappa$, where $c_V$ is the heat capacity, 
increases with temperature for the narrowest bridges. 

As mentioned above, to explain theoretically all these observations 
we have to work with the proper set of phonon modes. 
For wide membranes with parallel surfaces, the eigenmodes of 
the elastic equations are called \textit{Lamb modes} and 
\textit{horizontal shear modes}, as explained
for example in Ref.\ [\onlinecite{Auld:book}].
Using these modes and their 
dispersion relations, we could describe quite well the thermal properties 
of the membranes in the low temperature limit,\cite{PhysRevB.70.125425} 
i.e.\ at temperatures where the the characteristic thermal wavelength 
of the phonons is much longer than the thickness of the membrane. 

Nevertheless, the same temperature 
dependence of the heat conductivity and heat capacity persists also in a 
temperature range where this characteristic thermal wavelength becomes 
equal, or even smaller than the thickness. This can no longer be explained 
only by the dispersion relation of the Lamb modes in the membrane 
and we have to take into account the amorphous structure of the 
membrane and the resulting phonon scattering.

An amorphous material contains dynamic defects which can be modeled by 
an ensemble of two-level systems
(TLS).\cite{anderson:philmag,philips:lowtemp,esquinazi:book} A TLS  
can be understood as an atom 
or group of atoms which can tunnel between two close minima in the 
configuration space. Any deformation of the material disturbs the TLS, 
which can have a transition (an excitation or a de-excitation) to the other 
energy level. A passing phonon produces such a deformation and therefore 
may be scattered by the TLS. In bulk materials, the phonon modes are 
simple, transversally or longitudinally polarized plane waves and the 
deformation field they produce can be described by only two parameters,
the wave vector (or the wavelength) and the polarization. 
As a consequence, in the so called \textit{standard tunneling model}, the 
interaction Hamiltonian has a very simple
structure.\cite{anderson:philmag,philips:lowtemp,esquinazi:book}  
In a mesoscopic system, the deformation caused by the 
\textit{displacement field} 
of a phonon mode is more complex 
and the TLS-phonon interaction Hamiltonian has to be modified accordingly. 
This was done in Ref.\ [\onlinecite{PhysRevB.75.064202}]. Here 
we use this more general Hamiltonian to calculate the interaction of 
the phonon modes of the membrane with the TLSs. 

\section{TLS-phonon interaction} \label{sec_TLSph_int}

A TLS is described by a Hamiltonian which has the form 
\begin{eqnarray}
H_{\text{TLS}}
      &=& \frac{\Delta}{2}\sigma_z -\frac{\Lambda}{2}\sigma_x
\,, \label{eqn_TLS_Hamiltonian} 
\end{eqnarray}
when written in the basis formed by the 
ground states of the two potential wells between which
the system tunnels \cite{PhysRevB.75.064202}. 
In Eq.\ (\ref{eqn_TLS_Hamiltonian}) $\Delta$ is the 
\textit{asymmetry of the potential}, $\Lambda$ is the 
\textit{tunnel splitting}, and $\sigma_{x}$, $\sigma_{z}$ are Pauli 
matrices. The Hamiltonian 
(\ref{eqn_TLS_Hamiltonian}) can be diagonalized by an orthogonal 
transformation $O$, 
\begin{eqnarray}
H'_{\text{TLS}} &\equiv& O^T H_{\text{TLS}} O 
 = \frac{\epsilon}{2}\sigma_z \,,\label{eqn_TLS_Hamiltonian_diag} 
\end{eqnarray}
and we obtain the excitation energy, 
$\epsilon\equiv\sqrt{\Delta^2+\Lambda^2}$. Everywhere in this paper 
the superscript $T$ denotes the \textit{transpose} of a matrix. 

The TLS parameters $\Delta$ and $\Lambda$ are not the same for all 
the TLSs in the material, but they are
well modeled by the distribution 
$P(\Lambda,\Delta)=P_0/\Lambda$ in the unit volume of the material. 
We can rewrite the function $P$ in terms of the more practical variables
$\epsilon$ and $u\equiv\Lambda/\epsilon$,
\begin{equation}
P(\epsilon,u) = \frac{P_0}{u\sqrt{1-u^2}}\,. \label{eqn_TLS_density}
\end{equation}
The TLSs that have an excitation energy comparable to $k_{\text{B}}T$ are
very efficient phonon scatterers. 

The deformation due to the displacement field of a phonon is quantitatively 
described by the \textit{strain field}, which will be represented here 
by the 6-component vector $\bS$.\cite{Auld:book} If we denote the 
displacement field by $\bu(\br)$, then the strain is defined as 
the \textit{symmetric gradient} of $\bu(\br)$, 
i.e.\ $\bS^T\equiv(\nabla_S\bu)^T=(\partial_xu_x,\partial_yu_y,
\partial_zu_z,\partial_yu_z+\partial_zu_y,\partial_xu_z+\partial_zu_x,\partial_xu_y+\partial_yu_x)$. 
This deformation adds a 
time-dependent perturbation to the Hamiltonian (\ref{eqn_TLS_Hamiltonian}), 
which we shall denote by $H_1$. The perturbation is assumed to be diagonal, 
when written in the basis of the two ground states of the potential 
wells (like in Eq. (\ref{eqn_TLS_Hamiltonian})),
\cite{philips:lowtemp,anderson:philmag,ZPhys.257.212.1972,esquinazi:book,RevModPhys.59.1.Leggett} 
\begin{equation}
H_1 = \frac{\delta}{2}\sigma_z\,,\label{eqn_H_1_prime}
\end{equation}
and linear in the strain field at the location of the TLS,
\cite{RevModPhys.59.1.Leggett,PhysRevB.75.064202} 
\begin{equation}
\delta = 2\tilde\gamma\bT^t\cdot[r]\cdot\bS \,. \label{eqn_delta_do2}
\end{equation}
The other quantities in Eq.\ (\ref{eqn_delta_do2}) are the 
\textit{coupling constant} $\tilde\gamma$, the six component vector 
$\bT$, which is defined by a generic orientation $\hat\bt$ of the TLS as 
$\bT\equiv(t_x^2,t_y^2,t_z^2,2t_yt_z,2t_zt_x,2t_xt_y)^T$, and 
the $6\times 6$ matrix of the deformation potential parameters 
$[r]$. For isotropic materials, the matrix $[r]$ is 
\begin{eqnarray}
[r] &=& \left(\begin{array}{cccccc}
                1&\zeta&\zeta&0&0&0\\
                \zeta&1&\zeta&0&0&0\\
                \zeta&\zeta&1&0&0&0\\
                0&0&0&\xi&0&0\\
                0&0&0&0&\xi&0\\
                0&0&0&0&0&\xi\end{array}\right)\,,\label{eqn_r_def}
\end{eqnarray}
with the TLS potential parameters $\xi$ and $\zeta$ that satisfy 
the condition $\zeta+2\xi=1$.\cite{PhysRevB.75.064202} 

To calculate the scattering probabilities we have to write 
$H_1$ in the second quantization. For this we denote the TLS 
excited state by 
$\ket{\!\! \uparrow}$ and its ground state by $\ket{\!\! \downarrow}$ and we introduce 
the ``creation'' and ``annihilation'' operators $a^\dagger$ and $a$, 
respectively, so that $a^\dagger\ket{\!\! \downarrow}=\ket{\!\! \uparrow}$, 
$a\ket{\!\! \uparrow}=\ket{\!\! \downarrow}$, $a^\dagger\ket{\!\! \uparrow}=0$, and 
$a\ket{\!\! \downarrow}=0$. The operators $a$ and $a^\dagger$ obey Fermi 
commutation relations and in matrix form we have 
$\sigma_z=(2a^\dagger a-1)$ and $\sigma_x=(a^\dagger+a)$. 
The bosonic creation and annihilation operators for phonons will be denoted 
by $b_\mu^\dagger$ and $b_\mu$, respectively,
where $\mu$ stands in general for the quantum numbers of the phonon modes 
(see for example Refs.\ [\onlinecite{PhysRevB.75.064202,submitted.quantization}]). 
Using these notations and applying the transformation $O$ to the 
total Hamiltonian ($O^T(H_{\text{TLS}}+H_1)O\equiv H'_{\text{TLS}}+H'_1$), we obtain 
\begin{eqnarray}
H'_1
 &=&\frac{\tilde{\gamma}\Delta}{\epsilon}\mathbf{T}^T\cdot[r]\cdot
         \sum_\mu\left[\mathbf{S}_\mu b_\mu
                      +\mathbf{S}^\star_\mu b^\dagger_\mu \right](2a^\dagger a-1)
    \nonumber\\
 && \hspace{-2.5mm}
  -\frac{\tilde{\gamma}\Lambda}{\epsilon}\mathbf{T}^T\cdot[r]\cdot
         \sum_\mu\left[\mathbf{S}_\mu b_\mu
                      +\mathbf{S}^\star_\mu b^\dagger_\mu \right](a^\dagger+a)
    \,.\label{eqn_H_1_tilde}
\end{eqnarray}
In the first order perturbation theory, the phonon absorption and emission 
rates are determined by the off-diagonal elements of $H'_1$, i.e.\ of the 
second row of Eq.\ (\ref{eqn_H_1_tilde}). Higher order processes are not 
considered here. 

\subsection{The phonon modes in the membrane}

The phonon modes of a free standing, infinite membrane
are divided into three groups, according to their symmetry 
properties. One group is formed of simple, transversally polarized modes, 
called the horizontal shear modes ($h$).
The two other groups are the symmetric ($s$) and antisymmetric ($a$) 
Lamb modes. Together, these modes form a complete, 
orthonormal set of functions for the elastic displacement fields in the 
membrane and their proper quantization has been carried out in 
Ref.\ [\onlinecite{submitted.quantization}]. 
In this paper we shall use the results and notations from there and we shall 
call the three different types of phonons (i.e. $h$, $s$ and $a$)
\textit{polarizations}.

We assume that the membrane of thickness $d$ is placed parallel to the 
$(xy)$ plane and its parallel surfaces cut the $z$ axis at $\pm d/2$. 
The phonons propagate in the $(xy)$ plane with the wave vector 
$\bk_\parallel=k_\parallel\hat\bk_\parallel$, of real 
$k_\parallel$. We use \textit{hat} to denote unit 
vectors. 

Along the $z$ direction, the phonon modes are stationary. 
As the $h$ modes are pure transversal waves, they have only one 
wave vector component along the $z$ direction, which we denote by $k_h$.
The $s$ and $a$ waves are superpositions of transversal 
and longitudinal waves of wave vector components along the $z$ direction 
denoted by $k_t$ and $k_l$, respectively. 
Due to the boundary conditions, which demand that the membrane surfaces are 
stress-free, $k_h$ takes the discrete values $m\pi/d$, with $m$ taking all
integer values between $0$ and $\infty$,
whereas $k_t$ and $k_l$ satisfy the more complicated 
relations \cite{Auld:book} 
\begin{subequations}\label{eqn_disp}
\begin{eqnarray}
\frac{\tan(k_td/2)}{\tan(k_ld/2)}
 &=& -\frac{4k_tk_lk_\parallel^2}{(k_t^2-k_\parallel^2)^2}\,,\label{eqn_disp_sym}
\end{eqnarray}
for the symmetric modes and 
\begin{eqnarray}
\frac{\tan(k_ld/2)}{\tan(k_td/2)}
 &=& -\frac{4k_tk_lk_\parallel^2}{(k_t^2-k_\parallel^2)^2}\,.\label{eqn_disp_asym}
\end{eqnarray}
\end{subequations}
for the antisymmetric modes. Equations (\ref{eqn_disp}) plus Snell's law, 
$\omega^2=c_t^2(k_\parallel^2+k_t^2)=c_l^2(k_\parallel^2+k_l^2)$, enable
us to write $k_l$ as a function of $k_t$ for each of the polarizations 
$s$ and $a$. (In Snell's law, $c_t$ and $c_l$ are the respective
transversal and longitudinal sound velocities for bulk media.) 
Like in the case of the $h$ modes, the dispersion relations for the 
symmetric and antisymmetric modes split into branches, 
i.e.\ Eqs.\ (\ref{eqn_disp}) and Snell's law do not give only one 
function $k_l(k_t)$ for either set of modes, 
but produce an infinite, countable set of such functions. 
These functions will be called \textit{phonon branches} and we shall 
number them with $m=0,1,\ldots$, as we did in 
Ref.\ [\onlinecite{submitted.quantization}], where branches of bigger $m$ lie 
above branches of smaller $m$.

A simple way to express the quantum numbers of the phonon modes is to 
use Eqs.\ (\ref{eqn_disp}) and Snell's law to write the functions 
$k_t(k_\parallel)$ and $k_l(k_\parallel)$. Then each branch, of polarization 
$\sigma=s,a$ and branch number $m$, is going 
to be described by the continuous set of numbers 
$[k_l(k_\parallel),k_t(k_\parallel)]_{\sigma,m}$, with $k_\parallel$ 
taking values from 0 to $\infty$. We therefore choose the set $\mu$ of
quantum numbers that specify the phonon modes in Eq.\ (\ref{eqn_H_1_tilde}) to 
be $\mu\equiv\{\sigma,m,\bk_\parallel\}$. 

The functions $k_t(k_\parallel)$ and $k_l(k_\parallel)$ may take both, 
real and imaginary values. To distinguish between these situations, 
we write the imaginary values of $k_t$ as $i\kappa_t$ and the imaginary
values of $k_l$ as $i\kappa_l$. In these notations, $k_{l,t}$ and 
$\kappa_{l,t}$ take always positive, real values. 

In order to simplify the later discussion, we shall replace
$k_t$ and $k_l$ with the complex quantities
$\bar{k}_t\equiv k_t+i\kappa_t$ and $\bar{k}_l\equiv k_l+i\kappa_l$, 
respectively. Note, however, that $\bar{k}_t$ and $\bar{k}_l$ are 
never really complex, but they are \textit{either} 
real \textit{or} imaginary, as long as $k_\parallel$ is 
real.\cite{submitted.quantization}

The displacement fields of the phonon modes 
are\cite{submitted.quantization,Auld:book} 
\begin{widetext}
\begin{subequations}\label{eqn_displacements}
\begin{eqnarray}
\mathbf{u}_h
 &=& N_h\cos\left(k_h(z-d/2)\right)(\hat{\mathbf{k}}_\parallel\times\hat{\mathbf{z}})
     e^{i(\mathbf{k}_\parallel\cdot\mathbf{r}-\omega t)}\\
\nonumber\\
\mathbf{u}_s
 &=& N_s\left\{i\bar{k}_t\left(2k_\parallel^2\cos(\bar{k}_t d/2)\cos(\bar{k}_l z)\right.
     +[\bar{k}_t^2-k_\parallel^2]\cos(\bar{k}_l d/2)\cos(\bar{k}_t z)\right)\hat{\mathbf{k}}_\parallel
\nonumber\\
  && \hspace{5.5mm}-\left.k_\parallel\left(2\bar{k}_t\bar{k}_l\cos(\bar{k}_t d/2)\sin(\bar{k}_l z)
     -[\bar{k}_t^2-k_\parallel^2]\cos(\bar{k}_l d/2)\sin(\bar{k}_t z)\right)\hat{\mathbf{z}}\right\}
     e^{i(\mathbf{k}_\parallel\cdot\mathbf{r}-\omega t)}\\
\nonumber\\
\mathbf{u}_a
 &=& N_a\left\{i\bar{k}_t\left(2k_\parallel^2\sin(\bar{k}_t d/2)\sin(\bar{k}_l z)\right.
     +[\bar{k}_t^2-k_\parallel^2]\sin(\bar{k}_l d/2)\sin(\bar{k}_t z)\right)\hat{\mathbf{k}}_\parallel
\nonumber\\
  && \hspace{5.5mm}+\left.k_\parallel\left(2\bar{k}_t\bar{k}_l\sin(\bar{k}_t d/2)\cos(\bar{k}_l z)
     -[\bar{k}_t^2-k_\parallel^2]\sin(\bar{k}_l d/2)\cos(\bar{k}_t z)\right)\hat{\mathbf{z}}\right\}
     e^{i(\mathbf{k}_\parallel\cdot\mathbf{r}-\omega t)}\,.
\end{eqnarray}
\end{subequations}
\end{widetext}
As one can see, when $\bar{k}_t$ or $\bar{k}_l$ take imaginary values, 
the trigonometric functions
in (\ref{eqn_displacements}) will switch into hyperbolic functions. 
The normalization constants $N_h$, $N_s$ and $N_a$ are given by 
\cite{submitted.quantization}
\begin{widetext}
\begin{subequations}\label{eqn_N_complex}
\begin{eqnarray}
N_h^{-2}
 &=& \left\{\begin{array}{ll}
      V & m=0\\
      V/2 & m>0
     \end{array}\right.\label{eqn_N_h}\\N_s^{-2}
 &=& A \left\{
     4|\bar{k}_t|^2k_\parallel^2|\cos(\bar{k}_td/2)|^2\left(
     (|\bar{k}_l|^2+k_\parallel^2)\frac{\sinh(\kappa_ld)}{2\kappa_l}
    -(|\bar{k}_l|^2-k_\parallel^2)\frac{\sin(k_ld)}{2k_l}\right)\right.\nonumber\\
  && +|\bar{k}_t^2-k_\parallel^2|^2|\cos(\bar{k}_ld/2)|^2\left(
     (|\bar{k}_t|^2+k_\parallel^2)\frac{\sinh(\kappa_td)}{2\kappa_t}
    +(|\bar{k}_t|^2-k_\parallel^2)\frac{\sin(k_td)}{2k_t}\right)\nonumber\\
  &&-4k_\parallel^2|\cos(\bar{k}_ld/2)|^2\left(\kappa_t(|\bar{k}_t|^2+k_\parallel^2)
     \sinh(\kappa_td)-k_t(|\bar{k}_t|^2-k_\parallel^2)\sin(k_td)\right)
    \left.\vphantom{\frac{1}{2}}\right\}\label{eqn_N_s_complex}\\
N_a^{-2}
 &=& A \left\{
     4|\bar{k}_t|^2k_\parallel^2|\sin(\bar{k}_td/2)|^2\left(
     (|\bar{k}_l|^2+k_\parallel^2)\frac{\sinh(\kappa_ld)}{2\kappa_l}
    +(|\bar{k}_l|^2-k_\parallel^2)\frac{\sin(k_ld)}{2k_l}\right)\right.\nonumber\\
  && +|\bar{k}_t^2-k_\parallel^2|^2|\sin(\bar{k}_ld/2)|^2\left(
     (|\bar{k}_t|^2+k_\parallel^2)\frac{\sinh(\kappa_td)}{2\kappa_t}
    -(|\bar{k}_t|^2-k_\parallel^2)\frac{\sin(k_td)}{2k_t}\right)\nonumber\\
  && -4k_\parallel^2|\sin(\bar{k}_ld/2)|^2\left(\kappa_t(|\bar{k}_t|^2+k_\parallel^2)
     \sinh(\kappa_td)+k_t(|\bar{k}_t|^2-k_\parallel^2)\sin(k_td)\right)
     \left.\vphantom{\frac{1}{2}}\right\} \label{eqn_N_a_complex}\,,
\end{eqnarray}
\end{subequations}
\end{widetext}
where $A$ is the area of the membrane and $V=d\cdot A$ is the volume. To obtain the expressions 
for $N_s$ and $N_a$ for the different combinations of real or imaginary $\bar{k}_t$ and 
$\bar{k}_l$, one has to \textit{take the limit to} 0 of their redundant components in 
(\ref{eqn_N_s_complex}) and (\ref{eqn_N_a_complex}). For example if $\bar{k}_t$ is real and 
$\bar{k}_l$ is imaginary, we calculate the corresponding normalization factor by taking in 
Eq.\ (\ref{eqn_N_s_complex}) or (\ref{eqn_N_a_complex}) the limits $\kappa_t\to 0$ and $k_l\to 0$.
\cite{submitted.quantization} 
%
\section{Transition rates} \label{sec_transitions}

Now we have all the ingredients to calculate TLS transition rates or 
phonon absorption and emission probabilities. We shall denote the 
phonon-TLS quantum states by $\ket{n_\mu,\downarrow}$ or 
$\ket{n_\mu,\uparrow}$, where we denoted the number of 
phonons on the mode $\mu$ by $n_\mu$. Using Eq.\ (\ref{eqn_H_1_tilde}) 
we write the emission amplitude of a phonon by a TLS as 
\begin{eqnarray}
\bracket{n_\mu,\uparrow}{\tilde{H}_1|n_\mu+1,\downarrow}
 &=& -\frac{\tilde\gamma\Lambda}{\epsilon}\sqrt{\frac{\hbar n_\mu}{2\rho
\omega_\mu}}
      M_\mu\,,
\label{eqn_matrix_element}
\end{eqnarray}
with $M_\mu$ given by
\[
M_\mu(\hat\bt) = \mathbf{T}^T\cdot[r]\cdot\mathbf{S}_\mu \,.
\]
Explicitly, for the three phonon polarizations we have 
\begin{widetext}
\begin{subequations} \label{eqn_M_gen}
\begin{eqnarray}
M_{h,m,k_\parallel}(\hat\bt)
 &=& 2\xi N_h \left\{-t_yt_zk_h\sin(k_h(z-d/2)) 
    +it_xt_y k_\parallel\cos(k_h(z+d/2))\right\}e^{ik_\parallel x} \label{eqn_M_h_m_kpar} \\
M_{s,m,k_\parallel}(\hat\bt)
 &=& N_s\left\{-2\bar{k}_tk_\parallel\cos(\bar{k}_td/2)\cos(\bar{k}_lz)\{
     k_\parallel^2 [\zeta+(1-\zeta)t_x^2] + \bar{k}_l^2[\zeta+(1-\zeta)t_z^2]\}
     \right. \nonumber \\
  && + \bar{k}_tk_\parallel[\bar{k}_t^2-k_\parallel^2]\cos(\bar{k}_l
     d/2)\cos(\bar{k}_tz) (\zeta-1)(t_x^2-t_z^2) \nonumber \\
  && \left. -8it_xt_z\xi \bar{k}_t\bar{k}_l k_\parallel^2 
     \cos(\bar{k}_td/2)
     \sin(\bar{k}_lz) -2it_xt_z\xi[\bar{k}_t^2-k_\parallel^2]^2 
      \cos(\bar{k}_ld/2)\sin(\bar{k}_tz) \right\} \label{eqn_M_s_m_kpar} \\
M_{a,m,k_\parallel}(\hat\bt)
 &=& N_a\left\{-2\bar{k}_tk_\parallel \sin(\bar{k}_td/2)\sin(\bar{k}_lz) 
     \{k_\parallel^2 [t_x^2+\zeta(1-t_x^2)]  + \bar{k}_l^2 
     [t_z^2+\zeta(1-t_z^2)] \} \right. \nonumber \\  
 && - k_\parallel \bar{k}_t[\bar{k}_t^2-k_\parallel^2] 
    \sin(\bar{k}_ld/2) \sin(\bar{k}_tz)(1-\zeta)(t_x^2-t_z^2) \nonumber \\
 && \left.+ 8it_xt_z\xi \bar{k}_t\bar{k}_lk_\parallel^2 \sin(\bar{k}_td/2)
    \cos(\bar{k}_lz) +2t_xt_z\xi[\bar{k}_t^2-k_\parallel^2]^2 
     \sin(\bar{k}_ld/2) \cos(\bar{k}_tz) \right\} \label{eqn_M_a_m_kpar} 
\end{eqnarray}
\end{subequations}
\end{widetext}
where $k_t$ and $k_l$ are implicitly determined by the branch number, 
$m$, and $k_\parallel$. 
Using Fermi's golden rule, we calculate the phonon absorption and 
emission rates $\Gamma_{\text{abs}}^\mu$ and $\Gamma_{\text{em}}^\mu$, respectively,
\begin{subequations} \label{eqn_Gamma}
\begin{eqnarray}
\Gamma_{\text{abs}}^\mu
 &=& \frac{\pi}{\rho\omega_\mu}\frac{\tilde\gamma^2\Lambda^2}{\epsilon^2}
     |M_\mu|^2n_\mu
     \delta(\hbar\omega_\mu-\epsilon)\label{eqn_Gamma_abs}\\
\Gamma_{\text{em}}^\mu
 &=& \frac{\pi}{\rho\omega_\mu}\frac{\tilde\gamma^2\Lambda^2}{\epsilon^2}
     |M_\mu|^2(n_\mu+1)
     \delta(\hbar\omega_\mu-\epsilon)\label{eqn_Gamma_em}\,,
\end{eqnarray}
\end{subequations}
where $\epsilon$ is the energy of the TLS, as defined in Section 
\ref{sec_TLSph_int}, and $\omega$ is the angular frequency of the phonon. 

In an amorphous solid, the orientations of the TLSs are arbitrary, so the 
relevant quantities for our calculations are the averages of 
$\Gamma_{\text{abs}}^\mu$ over the directions $\hat\bt$ of the TLSs. 
The only quantity that depends on $\hat\bt$ in the Eqs.\ (\ref{eqn_Gamma}) 
is $|M_\mu|^2$. 
Additionally, we assume that the distribution of TLSs in the 
membrane is uniform, which leaves again $|M_\mu|^2$ the only quantity 
dependent on $z$ in the expressions for the absorbtion and emission rates. 
As we are interested in an average scattering probability rather than 
in a detailed description of where along the $z$ direction the scattering 
takes place, we also average $|M_\mu|^2$ along $z$. Denoting by 
$\langle\cdot\rangle$ 
the average over the TLS orientations and the $z$ variable, we obtain 
\begin{widetext}
\begin{subequations} \label{eqn_M2_average}
\begin{eqnarray}
\langle\left|M_h\right|^2\rangle
 &=& \frac{C_t}{V}(k_\parallel^2+k_h^2)\label{eqn_M_h}\\
\nonumber\\
\langle\left|M_s\right|^2\rangle
 &=& \frac{N_s^2}{d}\left\{4C_l|\bar{k}_t|^2k_\parallel^2|k_\parallel^2+\bar{k}_l^2|^2
     |\cos(\bar{k}_td/2)|^2
     \left(\frac{\sinh(\kappa_ld)}{2\kappa_l}+\frac{\sin(k_ld)}{2k_l}\right)\right.\nonumber\\
  && +C_t|\bar{k}_t^2-k_\parallel^2|^2|\cos(\bar{k}_ld/2)|^2\left((|\bar{k}_t|^2+k_\parallel^2)^2
     \frac{\sinh(\kappa_td)}{2\kappa_t}
    -(|\bar{k}_t|^2-k_\parallel^2)^2\frac{\sin(k_td)}{2k_t}\right)\nonumber\\
  && -2C_tk_t(2k_\parallel^6-k_\parallel^4|\bar{k}_t|^2
     -|\bar{k}_t|^6)|\cos(\bar{k}_ld/2)|^2\sin(k_td)\nonumber\\
  && -2C_t\kappa_t(2k_\parallel^6+k_\parallel^4|\bar{k}_t|^2+|\bar{k}_t|^6)|
     \cos(\bar{k}_ld/2)|^2\sinh(\kappa_td)\left.\vphantom{\frac{1}{1}}\right\}\label{eqn_M_s}\\
\nonumber\\
\langle\left|M_a\right|^2\rangle
 &=& \frac{N_a^2}{d}\left\{4C_l|\bar{k}_t|^2k_\parallel^2|k_\parallel^2+\bar{k}_l^2|^2
     |\sin(\bar{k}_td/2)|^2
     \left(\frac{\sinh(\kappa_ld)}{2\kappa_l}-\frac{\sin(k_ld)}{2k_l}\right)\right.\nonumber\\
  && +C_t|\bar{k}_t^2-k_\parallel^2|^2|\sin(\bar{k}_ld/2)|^2\left((|\bar{k}_t|^2+k_\parallel^2)
     \frac{\sinh(\kappa_td)}{2\kappa_t}+(|\bar{k}_t|^2-k_\parallel^2)\frac{\sin(k_td)}{2k_t}\right)\nonumber\\
  && -2C_tk_t(-2k_\parallel^6+k_\parallel^4|\bar{k}_t|^2
     +|\bar{k}_t|^6)|\sin(\bar{k}_ld/2)|^2\sin(k_td)\nonumber\\
  && -2C_t\kappa_t(2k_\parallel^6+k_\parallel^4|\bar{k}_t|^2
     +|\bar{k}_t|^6)|\sin(\bar{k}_ld/2)|^2\sinh(\kappa_td)
     \left.\vphantom{\frac{1}{1}}\right\}\,.\label{eqn_M_a}
\end{eqnarray}
\end{subequations}
\end{widetext}
$C_t$ and $C_l$ are constants that depend on the deformation potential
parameter $\xi$, $C_t=4\xi^2/15$ and $C_l=(15-40\xi+32\xi^2)/15$. 
Note that $C_l>C_t\ge 0$ for all $\xi$.\cite{PhysRevB.75.064202}
Using Eqs.\ (\ref{eqn_Gamma}) and (\ref{eqn_M2_average}), we can calculate the TLS
and phonon relaxation times,
\begin{equation}
\tau_\epsilon^{-1}
= \frac{\pi}{\rho}\frac{\tilde{\gamma}^2\Lambda^2}{\epsilon^2}
     \coth(\beta\epsilon/2)\sum_\mu\frac{1}{\omega_\mu}\langle|M_\mu|^2\rangle
     \delta(\hbar\omega_\mu-\epsilon) \label{eqn_tau_epsilon}
\end{equation}
and
\begin{eqnarray}
\tau_\mu^{-1}
 &=& \frac{\pi}{\rho}\frac{\tilde{\gamma}^2}{\omega_\mu}\sum_{\epsilon,u}u^2
     \langle|M_\mu|^2\rangle\tanh(\beta\epsilon/2)\delta(\hbar\omega_\mu-\epsilon)\nonumber\\
 &=& \frac{\pi\tilde{\gamma}^2VP_0}{\rho\omega_\mu}\langle|M_\mu|^2\rangle
     \tanh(\beta\hbar\omega/2) \,, \label{eqn_tau_mu}
\end{eqnarray}
respectively.
Here we changed the summation over $u$ and $\epsilon$ into a two-dimensional 
integral and used the TLS density (\ref{eqn_TLS_density}). 

Using Eq.\ (\ref{eqn_tau_mu}) we can calculate the two-dimensional heat conductivity 
along the membrane,
\begin{eqnarray}
\kappa 
 &=& \frac{1}{A}\sum_\mu\hbar\omega_\mu\tau_\mu (v_\mu)_x^2 
     \frac{\partial n_\mu}{\partial T} \label{eqn_kappa_T} \\
 &=& \frac{\hbar^2\rho}{16\pi^2\tilde{\gamma}^2VP_0}
     \frac{1}{k_{\text{B}}T^2} \nonumber\\
 &&  \times\sum_{n,\sigma}
     \int\limits_0^\infty dk_\parallel\left(\frac{\partial\omega_\mu}{\partial k_\parallel}\right)^2
     \frac{k_\parallel\omega_\mu^3}{\langle|M_\mu|^2\rangle}
     \frac{\coth(\beta\hbar\omega/2)}{\sinh^2(\beta\hbar\omega/2)}\, . \nonumber 
\end{eqnarray}
At an arbitrary temperature, $\kappa$ has to be calculated numerically. 
In this paper we give only the analytical low temperature approximation. 
%
\section{Low energy expansion: asymptotic results} \label{sec_asympt}

We can analytically calculate the scattering times or the thermal 
properties of the membrane only in the long wavelength limit, i.e.\ for
the branch $m=0$ for each of the three polarizations of the phonon 
modes and  $k_\parallel\ll 1/d$. The calculation of thermal properties 
in this limit corresponds to a temperature range in which 
$k_{\text{B}} T\ll \hbar c_t/d$. 

First we have to calculate the relaxation times for each polarization. 

%
\subsection{$h$ mode}
For the lowest branch of the $h$ mode, $\omega_{h,0}$ 
is linear in $k_\parallel$ and using Eq.\ (\ref{eqn_M_h}), the 
calculation of $\tau_{h,0,k_\parallel}$ is straightforward for 
any $k_\parallel$,
\begin{eqnarray}
\tau_{h,0,k_\parallel}
 &=& \frac{\hbar\rho c_t^2}{\pi\tilde{\gamma}^2P_0}
     \frac{1}{C_t}\frac{\coth(\beta\hbar\omega/2)}{\hbar\omega}\,. \label{tauh0}
\end{eqnarray}
%

%
\subsection{$s$ mode}
To get a long-wavelength expression of the dispersion relation 
for the lowest branch of the $s$ mode, we note that 
$\bar{k}_l=i\kappa_l$ takes imaginary values, which turns Eq.\ (\ref{eqn_disp_sym}) into
\begin{eqnarray}
\frac{\tan(k_tb/2)}{\tanh(\kappa_lb/2)}
 &=& \frac{4k_t\kappa_lk_\parallel^2}{(k_t^2-k_\parallel^2)^2}\,.\label{eqn_lowest_disp_sym}
\end{eqnarray}
We expand the trigonometric functions in Eq.\ (\ref{eqn_lowest_disp_sym}) to leading order
and obtain
\begin{equation}
\omega_{s,0,k_\parallel} = 2\frac{c_t}{c_l}\sqrt{c_l^2-c_t^2}k_\parallel\equiv 
c_sk_\parallel\,. \label{eqn_disp_s_01}
\end{equation}
Using this, we calculate the relaxation time for this branch, 
\begin{eqnarray}
\tau_{s,0,k_\parallel}
 &=& \frac{\hbar\rho c_t^2}{\pi\tilde{\gamma}^2P_0}
     \frac{4c_l^2(c_l^2-c_t^2)}{C_lc_t^4+C_tc_l^2(c_l^2-2c_t^2)}
     \frac{\coth(\beta\hbar\omega/2)}{\hbar\omega} \nonumber \\
 &\equiv&
     \frac{\hbar\rho c_t^2}{\pi\tilde{\gamma}^2P_0}
     \frac{1}{C_s}
     \frac{\coth(\beta\hbar\omega/2)}{\hbar\omega}\,. \label{eqn_taus0}
\end{eqnarray}
%
%
\subsection{$a$ mode}

The antisymmetric modes have a more complicated asymptotic expansion. 
First let us remark that for the lowest branch and any $k_\parallel$, 
both $\bar{k}_l=i\kappa_l$ and $\bar{k}_t=i\kappa_t$ take imaginary values, so we 
write Eq.\ (\ref{eqn_disp_asym}) in the form 
\begin{eqnarray}
\frac{\tanh(\kappa_ld/2)}{\tanh(\kappa_td/2)}
 &=& \frac{4\kappa_t\kappa_lk_\parallel^2}{(\kappa_t^2+k_\parallel^2)^2}\,.
     \label{eqn_lowest_disp_asym}
\end{eqnarray}
Expanding this equation to the second leading order, we obtain a quadratic 
dispersion relation for very small $k_\parallel$,\cite{PhysRevB.70.125425} 
\begin{eqnarray}
\omega_{a,0,k_\parallel} &=& dc_t\sqrt{\frac{c_l^2-c_t^2}{3c_l^2}}k_\parallel^2
\equiv \frac{\hbar}{2m^\star}k_\parallel^2\,. \label{eqn_disp_a_01}
\end{eqnarray}
Nevertheless, this asymptotic expression is not enough for 
the calculation of $\langle|M_{a,0,k_\parallel}|^2\rangle$, as it turns out 
that both expression (\ref{eqn_N_a_complex}) and expression (\ref{eqn_M_a}) 
are zero in the first and the second leading orders. Therefore, 
we have to expand to the third leading order to get non-zero 
results. From Eq.\ (\ref{eqn_disp_asym}) we obtain
\begin{eqnarray}
\omega_{a,0,k_\parallel}
 &=& \frac{\hbar}{2m^\star}\left(k_\parallel^2
     -d^2\frac{27c_l^2-20c_t^2}{90c_l^2}k_\parallel^4\right)\,,\label{eqn_disp_a_02}
\end{eqnarray}
from which we finally get
\begin{eqnarray}
\tau_{a,0,k_\parallel}
 &=& \frac{\hbar\rho c_t^2}{\pi\tilde{\gamma}^2P_0}
           \frac{c_l^2(c_l^2-c_t^2)}{C_lc_t^4+C_tc_l^2(c_l^2-2c_t^2)}
           \frac{\coth(\beta\hbar\omega/2)}{\hbar\omega} \nonumber \\
 &=& \frac{\hbar\rho c_t^2}{\pi\tilde{\gamma}^2P_0}
           \frac{1}{C_a}
           \frac{\coth(\beta\hbar\omega/2)}{\hbar\omega} \,, \label{eqn_tau_a_01} 
\end{eqnarray}
with $C_a=4C_s$.
\subsection{Comparison of the scattering rates and mean free paths}

The first thing to observe is that, although the dispersion relations 
for the $s$ and $a$ 
modes are different in the low $k_\parallel$ limit, 
$\tau_{s,0,k_\parallel}$ and $\tau_{a,0,k_\parallel}$ are related by the 
simple equation $\tau_{s,0,k_\parallel}=4\tau_{a,0,k_\parallel}$. In other words, 
the scattering rate for the $s$ phonons is 4 times smaller 
than the scattering rate of $a$ phonons at the same $\omega$. 

Let us now compare $\tau_{a,0,k_\parallel}$ with $\tau_{h,0,k_\parallel}$.
For this, we calculate the ratio 
\begin{eqnarray}
\frac{\tau_{h,0,k_\parallel}}{\tau_{a,0,k_\parallel}}
 &=&      \frac{C_lc_t^4+C_tc_l^2(c_l^2-2c_t^2)}{C_tc_l^2(c_l^2-c_t^2)} \nonumber \\
 &=&      1-(c_t/c_l)^2\frac{1-(C_l/C_t)(c_t/c_l)^2}{1- (c_t/c_l)^2}
\label{eqn_tau_h_over_tau_a}
\end{eqnarray}
\begin{figure}
\begin{center}
\resizebox{70mm}{!}{\includegraphics{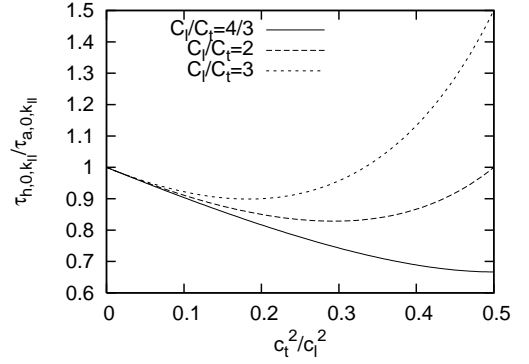}}
\caption{The ratio $\tau_{h,0,k_\parallel}/\tau_{a,0,k_\parallel}$ 
as function of the ratio $c_t^2/c_l^2$ for different values
of $C_l/C_t$. The value $C_l/C_t=4/3$ is the minimum possible value
and thus all other curves lie above the solid curve in the plot.}
\label{fig_tau_ratio}
\end{center}
\end{figure}
In any normal material (i.e.\ with positive Poisson ratio), the ratio 
$c_t^2/c_l^2$ is restricted to $0<c_t^2/c_l^2\le 1/2$ and for $C_l/C_t$ we have
\cite{PhysRevB.75.064202} $C_l/C_t=15/4\xi^2-10/\xi+8\ge 4/3$. 
In Fig.\ \ref{fig_tau_ratio}, we plot the ratio 
$\tau_{h,0,k_\parallel}/\tau_{a,0,k_\parallel}$ as function of 
$c_t^2/c_l^2$ for different values of $C_l/C_t$.

We first remark that in the limit $c_t/c_l\to 0$, 
$\tau_{h,0,k_\parallel}/\tau_{a,0,k_\parallel}=1$, 
independent of the value for $C_l/C_t$. Increasing $c_t/c_l$ will result in a
decrease of $\tau_{h,0,k_\parallel}/\tau_{a,0,k_\parallel}$, until a minimum is reached
at $(c_t/c_l)^2=1-\sqrt{1-(C_l/C_t)^{-1}}$, which lies between $0$ and $1/2$. Afterwards 
$\tau_{h,0,k_\parallel}/\tau_{a,0,k_\parallel}$ increases monotonically until it
reaches the value $(C_l/C_t)/2$ for $(c_t/c_l)^2=1/2$. As 
$\tau_{h,0,k_\parallel}/\tau_{a,0,k_\parallel}$ increases monotonically with 
$C_l/C_t$, we conclude that $\tau_{h,0,k_\parallel}<\tau_{a,0,k_\parallel}$ for any $c_t/c_l$, 
as long as $C_l/C_t < 2$. For $C_l/C_t\ge 2$, 
$\tau_{h,0,k_\parallel}$ can be either smaller or greater than $\tau_{a,0,k_\parallel}$,
depending on whether $(c_t/c_l)^2$ is smaller or greater than $C_t/C_l$, respectively.
A typical value for $(c_t/c_l)^2$ in SiN$_x$ is $0.36$, which means that 
$\tau_{h,0,k_\parallel}<\tau_{a,0,k_\parallel}$ as long as $C_l/C_t$ is smaller than
$2.78$.

When comparing $\tau_{s,0,k_\parallel}$ with $\tau_{h,0,k_\parallel}$, we encounter a
similar situation. As $\tau_{s,0,k_\parallel}=4\tau_{a,0,k_\parallel}$ the ratio
$\tau_{h,0,k_\parallel}/\tau_{s,0,k_\parallel}$ has the same features as the
ratio $\tau_{h,0,k_\parallel}/\tau_{a,0,k_\parallel}$, except for the fact that
the critical value for $C_l/C_t$ is $8$, i.e.\ $\tau_{h,0,k_\parallel}$
is always smaller than $\tau_{s,0,k_\parallel}$, if 
$C_l/C_t<8$, and can be either smaller or greater than 
$\tau_{s,0,k_\parallel}$ for $C_l/C_t\ge 8$.
For the SiN$_x$ typical value, $(c_t/c_l)^2=0.36$, we have 
$\tau_{h,0,k_\parallel}\le\tau_{s,0,k_\parallel}$ as long as $C_l/C_t\le17.6$. 

More interesting than the scattering rates, is to compare the phonon 
mean free paths, since these can be directly measured experimentally. 
For this, let us first use the dispersion relations (\ref{eqn_disp_s_01}), 
(\ref{eqn_disp_a_01}), and $\omega_{h,0,k_\parallel}=c_t k_\parallel$ 
to write the expressions for the mean free paths: 
\begin{subequations} \label{eqn_path_h}
\begin{eqnarray}
l_{h,0,k_\parallel}
 &=& c_t\tau_{h,0,k_\parallel} \nonumber\\
 &=& \frac{\hbar\rho c_t^3}{\pi\tilde{\gamma}^2P_0}
     \frac{1}{C_t}\frac{\coth(\beta\hbar\omega/2)}{\hbar\omega} 
     \label{eqn_path_h0_omega} \\ 
 &=& \frac{\hbar\rho c_t^2}{\pi\tilde{\gamma}^2P_0}
     \frac{1}{C_t}\frac{\coth(\beta\hbar c_t k_\parallel/2)}
     {\hbar k_\parallel} \,, \label{eqn_path_h0_kpar} 
\end{eqnarray}
\end{subequations}
\begin{subequations} \label{eqn_path_s}
\begin{eqnarray}
l_{s,0,k_\parallel}
 &=& c_s\tau_{s,0,k_\parallel} \nonumber\\
 &=& \frac{\hbar\rho c_t^3}{\pi\tilde{\gamma}^2P_0}
     \frac{2\sqrt{1-c_t^2/c_l^2}}{C_s} \frac{\coth(\beta\hbar\omega/2)}{\hbar\omega}
     \label{eqn_path_s0_omega} \\ 
&=& \frac{\hbar\rho c_t^2}{\pi\tilde{\gamma}^2P_0}
     \frac{1}{C_s}\frac{\coth(\beta\hbar c_s k_\parallel/2)}{\hbar k_\parallel}
 \,, \label{eqn_path_s0_kpar} 
\end{eqnarray}
\end{subequations}
\begin{subequations} \label{eqn_path_a}
\begin{eqnarray}
l_{a,0,k_\parallel}
 &=& \sqrt{\frac{2\hbar\omega}{m^\star}}\tau_{a,0,k_\parallel} \nonumber \\
 &=& \frac{\hbar\rho c_t^3}{\pi\tilde{\gamma}^2P_0}\frac{2(1-c_t^2/c_l^2)^{1/4}}{3^{1/4}C_a}
     \sqrt{\frac{d\omega}{c_t}}\nonumber\\
  && \times\frac{\coth(\beta\hbar\omega/2)}{\hbar\omega}\label{eqn_path_a0_omega} \\
 &=& \frac{\hbar\rho c_t^2}{\pi\tilde{\gamma}^2P_0}
     \frac{2}{C_a}\frac{\coth\left(\frac{\beta\hbar dc_t}{2\sqrt{3}} 
     \sqrt{1-c_t^2/c_l^2}k_\parallel^2\right)}{\hbar k_\parallel}
 \,\label{eqn_path_a0_kpar} 
\end{eqnarray}
\end{subequations}

A way to determine the mean free path of phonons is to 
measure the resonant attenuation of ultra-sound, propagating along the membrane. 
If this is experimentally impossible, another way to determine 
the material parameters is to make acoustic measurements on thicker and 
wider membranes. 
Note however that elastic waves attenuate not only because of resonant 
scattering of phonons (Eqs.\ \ref{eqn_path_h}-\ref{eqn_path_a}), 
but also due to energy relaxation.\cite{ZPhys.257.212.1972} 
Nevertheless, since we are interested here in the thermal properties 
of the membranes, only the resonant scattering is important and 
we disregard the energy relaxation mechanism.

To analyze the results (\ref{eqn_path_h}), (\ref{eqn_path_s}), 
and (\ref{eqn_path_a}) in more detail, we expressed the mean 
free paths both in terms of the angular frequency and in terms of 
$k_\parallel$. If the elastic modes of different polarizations are 
produced with the same $\omega$, then we should compare the mean free paths 
as given by the expressions (\ref{eqn_path_h0_omega}), (\ref{eqn_path_s0_omega}), 
and (\ref{eqn_path_a0_omega}), which we denote as $l_{\sigma,0,\omega}$. 
For example, $l_{h,0,\omega}/l_{s,0,\omega}=(\tau_{h,0,\omega}/\tau_{s,0,\omega})\cdot(c_t/c_s)$ 
which is smaller than $(\tau_{h,0,\omega}/\tau_{s,0,\omega})$ in any 
material. The discussion we made above about 
$\tau_{h,0,\omega}/\tau_{s,0,\omega}$ applies here too. 

The expressions (\ref{eqn_path_a}) for $l_{a,0,\omega}$ are very different 
from the ones for $l_{h,0,\omega}$ and $l_{s,0,\omega}$. 
Nevertheless, the expressions (\ref{eqn_path_h}), (\ref{eqn_path_s}), and 
(\ref{eqn_path_a}) are calculated for $k_\parallel\ll 1/d$, 
which means that $\omega d\ll (2\pi/k_\parallel)\omega=c_t$ which implies 
$\sqrt{\omega d/c_t}\ll 1$. 
Taking this into account when we compare the expressions 
for $l_{h,0,\omega}$, $l_{s,0,\omega}$, and $l_{a,0,\omega}$, at the same 
$\omega$, 
we conclude that, as a function of frequency, for low enough frequencies 
the antisymmetric Lamb modes 
have the shortest mean free path. This is a consequence of the fact that 
the group velocity of the $a$ modes decreases to zero as $k_\parallel$ 
decreases.

If we compare $l_{h,0,k_\parallel}$ and $l_{s,0,k_\parallel}$ as functions 
of $k_\parallel$, we see that 
$l_{h,0,k_\parallel}/l_{s,0,k_\parallel}=(\tau_{h,0,k_\parallel}
/\tau_{s,0,k_\parallel})\cdot[\coth(\beta\hbar c_t k_\parallel/2)
/\coth(\beta\hbar c_s k_\parallel/2)]$, which is bigger than 
$\tau_{h,0,k_\parallel}/\tau_{s,0,k_\parallel}$, since $c_t<c_s$ implies 
$\coth(\beta\hbar c_t k_\parallel/2)>\coth(\beta\hbar c_s k_\parallel/2)$. 
Comparing
 $l_{a,0,k_\parallel}$ with the expressions 
(\ref{eqn_path_h0_kpar}) and (\ref{eqn_path_s0_kpar}), we observe, for
example, that both
$l_{a,0,k_\parallel}/l_{h,0,k_\parallel}$ and
$l_{a,0,k_\parallel}/l_{s,0,k_\parallel}$ are proportional to 
$\coth(\beta\hbar\omega_{a,0,k_\parallel}/2)/\coth(\beta\hbar\omega_{h,0,k_\parallel}/2)$.
But again, for long wavelengths, due to the quadratic dependence of 
$\omega_{a,0,k_\parallel}$ on $k_\parallel$, we have 
$\omega_{a,0,k_\parallel}\ll\omega_{h/s,0,k_\parallel}$. Moreover, 
$\coth(x)\sim1/x$ for $x\to 0$, so both 
$l_{a,0,k_\parallel}/l_{h,0,k_\parallel}$ and $l_{a,0,k_\parallel}/l_{s,0,k_\parallel}$ are proportional to $1/k_\parallel$ and become very big 
in the limit of long wavelengths. 
In conclusion, as function of $k_\parallel$ in the limit $dk_\parallel\ll 1$, 
the antisymmetric modes have a much longer mean free path than the symmetric 
and the horizontal shear modes with the same $k_\parallel$.

\subsection{Calculation of the heat conductivity} \label{kappa_at_last}

Now we calculate the heat conductivity in the limit of low temperature. 
In that limit, only the lowest branch of each polarization will be occupied
and we can write Eq.\ (\ref{eqn_kappa_T}) in the form 
\begin{eqnarray}
\kappa 
 &=& \frac{\hbar^2}{16\pi}\frac{1}{k_{\text{B}}T^2}\sum_\sigma
\int\limits_{\omega^\star_{\sigma,0}}^\infty d\omega
     \frac{k_{\parallel,\sigma,0}(\omega)l_{\sigma,0,k_\parallel}\omega^2}
          {\sinh^2(\beta\hbar\omega/2)}\,,\label{eqn_conductivity_low_T}
\end{eqnarray}
where the lower limits $\omega^\star_{\sigma,0}$ were introduced for the
reasons that will become clear imediately. 
Using Eqs.\ (\ref{eqn_path_h}-\ref{eqn_path_a})
for the mean free paths we express $\kappa$ as a sum of three contributions:
\begin{equation}
\kappa = \frac{k_B^2\rho c_t^2}{16\pi^2\hbar\tilde{\gamma}^2P_0}T
     \left(\frac{I(x^\star_{h,0})}{C_t}+\frac{I(x^\star_{s,0})}{C_s}+\frac{2I(x^\star_{a,0})}
{C_a}\right) ,
     \label{eqn_kappa_final}
\end{equation}
where $x^\star_{\sigma,0}\equiv\beta\hbar\omega^\star_{\sigma,0}$ and by $I(x)$ we 
denoted the integral 
\begin{eqnarray}
I(x)
 &\equiv& \int_x^\infty dy\frac{y^2\coth(y/2)}{\sinh^2(y/2)} \nonumber\\
      &=& \frac{4x^2e^x}{(e^x-1)^2}+\frac{8x}{e^x-1}-8\ln(1-e^{-x}) .
\label{eqn_primitive}
\end{eqnarray}
Note that, although the mean free paths for the $h$ and $s$ 
modes have different functional dependences on $\omega$ than the mean 
free path for the $a$ modes, the 
integrand in Eq.\ (\ref{eqn_kappa_final}) is the same for all three modes. 
The role of the lower cut-off in Eq.\ (\ref{eqn_conductivity_low_T}) 
becomes obvious when we look at Eq.\ (\ref{eqn_primitive}): 
the integral $I(x)$ has a logarithmic divergence in $x=0$. 

If the cut-off is small enough, then we can approximate $I(x)$ by 
\begin{equation}
I(x) \approx 12-8\ln(x) \, \label{Ilowx}
\end{equation}
and inserting this into (\ref{eqn_kappa_final}) we obtain 
\begin{eqnarray}
\kappa &=& \frac{k_B^2\rho c_t^2}{4\pi^2\hbar\tilde{\gamma}^2P_0}T
     \left[\frac{3-2\ln(\beta\hbar\omega^\star_{h,0})}{C_t} \right.\nonumber \\
&& \left.+\frac{3-2\ln(\beta\hbar\omega^\star_{s,0})}{C_s}
+\frac{2[3-2\ln(\beta\hbar\omega^\star_{a,0})]}
{C_a}\right] ,
     \label{eqn_kappa_final1}
\end{eqnarray}
where the first, the second, and the third terms in the square brackets above 
give the contributions of the $h$, $s$, and $a$ phonon modes to the heat 
conductivity. The above expression leads to the temperature dependence 
$\kappa\propto T(a+b\ln T)$ and this dependence is a hallmark of the 
TLS-limited heat conductance at low temperature. 

For a numerical estimate let us use for the cut-off the finite size of the 
membrane, which limits the wave vectors to values of the order of 
$2\pi/\sqrt{A}$. 
For the typical experimental parameters $T=0.1$~K, 
$\sqrt{A}=400$~${\rm \mu m}$ and $d=200$~nm,\cite{Leivo:thesis,ApplPhysLett.72.1305.Leivo,PhysRevLett.81.2958.Anghel} we have 
$\ln(x_{h,0,2\pi/L})\approx -4.9$, $\ln(x_{s,0,2\pi/L})\approx -4.4$, 
and $\ln(x_{a,0,2\pi/L})\approx-11.4$. But since $C_a=4C_s$ 
(see Eqs.\ (\ref{eqn_taus0}) and (\ref{eqn_tau_a_01})), the contributions of all the 
phonon polarizations to the heat conductivity are of the same order. 

\section{Conclusions} \label{sec_concl}

We used the model introduced in Ref.\ [\onlinecite{PhysRevB.75.064202}] to 
calculate the scattering of the elastic modes in a thin, amorphous membrane. 
We modeled the scattering centers in the membrane by an ensemble of 
TLSs with the same properties and distribution over energy splitting and 
asymmetry as the TLSs in a bulk material. If this assumption is 
valid remains to be checked by experiment. We obtained the expressions 
for the TLS relaxation time (\ref{eqn_tau_epsilon}), for the phonon scattering 
time (\ref{eqn_tau_mu}) and for the heat conductivity $\kappa$ (\ref{eqn_kappa_T}). 

For general temperatures, the heat conductivity and the scattering times 
have to be calculated numerically. We calculated 
analytical 
low temperature approximations and compared the mean free paths of 
different phonon polarizations. 
In this way we observed that the contribution of the lowest branches of 
the phonon modes to the heat conductivity are logarithmically divergent at $k_\parallel\to 0$. 
This could be a reason for which in some experiments a 
radiative heat transport is observed.\cite{ApplPhysLett86.251903.Hoevers} 
Nevertheless, there is a natural lower cut-off of $k_\parallel\to 0$ due to the 
finite size of the membrane. This cut-off renders 
$\kappa$ finite, which, in the low temperature limit 
behaves like $\kappa\propto T(a+b\ln T)$. This behavior is a 
hallmark of the TLS-limited heat conductance at low temperature. 

Due to the dispersion relations of the phonon modes, the TLSs distribution 
in the low energy limit has a bigger impact on the heat conductivity 
in thin membranes than in bulk materials. If we for instance modify 
the distribution (\ref{eqn_TLS_density}) into 
\begin{eqnarray*}
P'(\epsilon,u) &=& \frac{P_0}{\epsilon^\alpha u\sqrt{1-u^2}}\,,
\end{eqnarray*}
with an extra energy dependence, $\epsilon^{-\alpha}$, we make the expression 
(\ref{eqn_conductivity_low_T}) for $\kappa$ convergent even in the 
$\omega_{\sigma,0}^\star\to 0$ limit, which leads to a low temperature asymptotic 
dependence of $\kappa\propto T^{1+\alpha}$. But if this is the situation or not
has to be decided experimentally. 

\acknowledgments 

Discussions with J. P. Pekola, I. J. Maasilta, and V. Vinokur are
gratefully  acknowledged. 
This work was partly supported  by the U. S.
Department of  Energy Office of Science  under the Contract No.
DE-AC02-06CH11357 and by the NATO grant EAP.RIG 982080. DVA acknowledges 
the hospitality of the University of Jyv\"askyl\"a, where part of this 
work has been carried-out, and the financial support from the Academy
of Finland.


\begin{thebibliography}{16}
\expandafter\ifx\csname natexlab\endcsname\relax\def\natexlab#1{#1}\fi
\expandafter\ifx\csname bibnamefont\endcsname\relax
  \def\bibnamefont#1{#1}\fi
\expandafter\ifx\csname bibfnamefont\endcsname\relax
  \def\bibfnamefont#1{#1}\fi
\expandafter\ifx\csname citenamefont\endcsname\relax
  \def\citenamefont#1{#1}\fi
\expandafter\ifx\csname url\endcsname\relax
  \def\url#1{\texttt{#1}}\fi
\expandafter\ifx\csname urlprefix\endcsname\relax\def\urlprefix{URL }\fi
\providecommand{\bibinfo}[2]{#2}
\providecommand{\eprint}[2][]{\url{#2}}

\bibitem[{\citenamefont{Leivo}(1999)}]{Leivo:thesis}
\bibinfo{author}{\bibfnamefont{M.}~\bibnamefont{Leivo}}, Ph.D. thesis,
  \bibinfo{school}{University of Jyv\"askyl\"a} (\bibinfo{year}{1999}).

\bibitem[{\citenamefont{Leivo and Pekola}(1998)}]{ApplPhysLett.72.1305.Leivo}
\bibinfo{author}{\bibfnamefont{M.~M.} \bibnamefont{Leivo}} \bibnamefont{and}
  \bibinfo{author}{\bibfnamefont{J.~P.} \bibnamefont{Pekola}},
  \bibinfo{journal}{Appl. Phys. Lett.} \textbf{\bibinfo{volume}{72}},
  \bibinfo{pages}{1305} (\bibinfo{year}{1998}).

\bibitem[{\citenamefont{Holmes et~al.}(1998)\citenamefont{Holmes, Gildemeister,
  Richards, and Kotsubo}}]{ApplPhysLett.72.2250.Holmes}
\bibinfo{author}{\bibfnamefont{W.}~\bibnamefont{Holmes}},
  \bibinfo{author}{\bibfnamefont{J.~M.} \bibnamefont{Gildemeister}},
  \bibinfo{author}{\bibfnamefont{P.~L.} \bibnamefont{Richards}},
  \bibnamefont{and} \bibinfo{author}{\bibfnamefont{V.}~\bibnamefont{Kotsubo}},
  \bibinfo{journal}{Appl. Phys. Lett.} \textbf{\bibinfo{volume}{72}},
  \bibinfo{pages}{2250} (\bibinfo{year}{1998}).

\bibitem[{\citenamefont{Woodcraft et~al.}(2000)\citenamefont{Woodcraft,
  Sudiwalaa, Wakui, Bhatia, Bock, and Turner}}]{PhysicaB.284.1968.Woodcraft}
\bibinfo{author}{\bibfnamefont{A.~L.} \bibnamefont{Woodcraft}},
  \bibinfo{author}{\bibfnamefont{R.~V.} \bibnamefont{Sudiwalaa}},
  \bibinfo{author}{\bibfnamefont{E.}~\bibnamefont{Wakui}},
  \bibinfo{author}{\bibfnamefont{R.~S.} \bibnamefont{Bhatia}},
  \bibinfo{author}{\bibfnamefont{J.~J.} \bibnamefont{Bock}}, \bibnamefont{and}
  \bibinfo{author}{\bibfnamefont{A.~D.} \bibnamefont{Turner}},
  \bibinfo{journal}{Physica B: Cond. Matt.} \textbf{\bibinfo{volume}{284}},
  \bibinfo{pages}{1968} (\bibinfo{year}{2000}).

\bibitem[{\citenamefont{Hoevers et~al.}(2005)\citenamefont{Hoevers, Ridder,
  Germeau, Bruijn, de~Korte, and Wiegerink}}]{ApplPhysLett86.251903.Hoevers}
\bibinfo{author}{\bibfnamefont{H.~F.~C.} \bibnamefont{Hoevers}},
  \bibinfo{author}{\bibfnamefont{M.~L.} \bibnamefont{Ridder}},
  \bibinfo{author}{\bibfnamefont{A.}~\bibnamefont{Germeau}},
  \bibinfo{author}{\bibfnamefont{M.~P.} \bibnamefont{Bruijn}},
  \bibinfo{author}{\bibfnamefont{P.~A.~J.} \bibnamefont{de~Korte}},
  \bibnamefont{and} \bibinfo{author}{\bibfnamefont{R.~J.}
  \bibnamefont{Wiegerink}}, \bibinfo{journal}{Appl. Phys. Lett.}
  \textbf{\bibinfo{volume}{86}}, \bibinfo{eid}{251903}
  (pages~\bibinfo{numpages}{3}) (\bibinfo{year}{2005}).

\bibitem[{\citenamefont{Zink and
  Hellman}(2004)}]{SolidStateComm.129.199.ZinkHellman}
\bibinfo{author}{\bibfnamefont{B.~L.} \bibnamefont{Zink}} \bibnamefont{and}
  \bibinfo{author}{\bibfnamefont{F.}~\bibnamefont{Hellman}},
  \bibinfo{journal}{Solid State Comm.} \textbf{\bibinfo{volume}{129}},
  \bibinfo{pages}{199} (\bibinfo{year}{2004}).

\bibitem[{\citenamefont{Anghel et~al.}(1998)\citenamefont{Anghel, Pekola,
  Leivo, Suoknuuti, and Manninen}}]{PhysRevLett.81.2958.Anghel}
\bibinfo{author}{\bibfnamefont{D.~V.} \bibnamefont{Anghel}},
  \bibinfo{author}{\bibfnamefont{J.~P.} \bibnamefont{Pekola}},
  \bibinfo{author}{\bibfnamefont{M.~M.} \bibnamefont{Leivo}},
  \bibinfo{author}{\bibfnamefont{J.~K.} \bibnamefont{Suoknuuti}},
  \bibnamefont{and} \bibinfo{author}{\bibfnamefont{M.}~\bibnamefont{Manninen}},
  \bibinfo{journal}{Phys. Rev. Lett.} \textbf{\bibinfo{volume}{81}},
  \bibinfo{pages}{2958} (\bibinfo{year}{1998}).

\bibitem[{\citenamefont{Auld}(1990)}]{Auld:book}
\bibinfo{author}{\bibfnamefont{B.~A.} \bibnamefont{Auld}},
  \emph{\bibinfo{title}{Acoustic Fields and Waves in Solids, 2nd Ed.}}
  (\bibinfo{publisher}{Robert E. Krieger Publishing Company},
  \bibinfo{year}{1990}), ISBN \bibinfo{isbn}{0-89874-783}.

\bibitem[{\citenamefont{K\"uhn et~al.}(2004)\citenamefont{K\"uhn, Anghel,
  Pekola, Manninen, and Galperin}}]{PhysRevB.70.125425}
\bibinfo{author}{\bibfnamefont{T.}~\bibnamefont{K\"uhn}},
  \bibinfo{author}{\bibfnamefont{D.~V.} \bibnamefont{Anghel}},
  \bibinfo{author}{\bibfnamefont{J.~P.} \bibnamefont{Pekola}},
  \bibinfo{author}{\bibfnamefont{M.}~\bibnamefont{Manninen}}, \bibnamefont{and}
  \bibinfo{author}{\bibfnamefont{Y.~M.} \bibnamefont{Galperin}},
  \bibinfo{journal}{Phys. Rev. B} \textbf{\bibinfo{volume}{70}},
  \bibinfo{pages}{125425} (\bibinfo{year}{2004}).

\bibitem[{\citenamefont{Anderson et~al.}(1972)\citenamefont{Anderson, Halperin,
  and Varma}}]{anderson:philmag}
\bibinfo{author}{\bibfnamefont{P.~W.} \bibnamefont{Anderson}},
  \bibinfo{author}{\bibfnamefont{B.~I.} \bibnamefont{Halperin}},
  \bibnamefont{and} \bibinfo{author}{\bibfnamefont{C.~M.} \bibnamefont{Varma}},
  \bibinfo{journal}{Phil. Mag.} \textbf{\bibinfo{volume}{25}},
  \bibinfo{pages}{1} (\bibinfo{year}{1972}).

\bibitem[{\citenamefont{Philips}(1972)}]{philips:lowtemp}
\bibinfo{author}{\bibfnamefont{W.~A.} \bibnamefont{Philips}},
  \bibinfo{journal}{J. Low Temp. Phys.} \textbf{\bibinfo{volume}{7}},
  \bibinfo{pages}{351} (\bibinfo{year}{1972}).

\bibitem[{\citenamefont{Esquinazi}(1998)}]{esquinazi:book}
\bibinfo{author}{\bibfnamefont{P.}~\bibnamefont{Esquinazi}},
  \emph{\bibinfo{title}{Tunneling systems in amorphous and crystalline solids}}
  (\bibinfo{publisher}{Springer}, \bibinfo{year}{1998}), ISBN
  \bibinfo{isbn}{3-540-63960-8}.

\bibitem[{\citenamefont{Anghel et~al.}(2007)\citenamefont{Anghel, K\"uhn,
  Galperin, and Manninen}}]{PhysRevB.75.064202}
\bibinfo{author}{\bibfnamefont{D.~V.} \bibnamefont{Anghel}},
  \bibinfo{author}{\bibfnamefont{T.}~\bibnamefont{K\"uhn}},
  \bibinfo{author}{\bibfnamefont{Y.~M.} \bibnamefont{Galperin}},
  \bibnamefont{and} \bibinfo{author}{\bibfnamefont{M.}~\bibnamefont{Manninen}},
  \bibinfo{journal}{Phys. Rev. B} \textbf{\bibinfo{volume}{75}},
  \bibinfo{pages}{064202} (\bibinfo{year}{2007}).

\bibitem[{\citenamefont{J\"ackle}(1972)}]{ZPhys.257.212.1972}
\bibinfo{author}{\bibfnamefont{J.}~\bibnamefont{J\"ackle}},
  \bibinfo{journal}{Z. Phys.} \textbf{\bibinfo{volume}{257}},
  \bibinfo{pages}{212} (\bibinfo{year}{1972}).

\bibitem[{\citenamefont{Leggett et~al.}(1987)\citenamefont{Leggett,
  Chakravarty, Dorsey, Fisher, Garg, and Zwerger}}]{RevModPhys.59.1.Leggett}
\bibinfo{author}{\bibfnamefont{A.~J.} \bibnamefont{Leggett}},
  \bibinfo{author}{\bibfnamefont{S.}~\bibnamefont{Chakravarty}},
  \bibinfo{author}{\bibfnamefont{A.~T.} \bibnamefont{Dorsey}},
  \bibinfo{author}{\bibfnamefont{M.~P.~A.} \bibnamefont{Fisher}},
  \bibinfo{author}{\bibfnamefont{A.}~\bibnamefont{Garg}}, \bibnamefont{and}
  \bibinfo{author}{\bibfnamefont{W.}~\bibnamefont{Zwerger}},
  \bibinfo{journal}{Rev. Mod. Phys.} \textbf{\bibinfo{volume}{59}},
  \bibinfo{pages}{1} (\bibinfo{year}{1987}).

\bibitem[{\citenamefont{Anghel and K\"uhn}(2006)}]{submitted.quantization}
\bibinfo{author}{\bibfnamefont{D.~V.} \bibnamefont{Anghel}} \bibnamefont{and}
  \bibinfo{author}{\bibfnamefont{T.}~\bibnamefont{K\"uhn}},
  \bibinfo{journal}{(submitted) cond-mat/0611528}  (\bibinfo{year}{2006}).

\end{thebibliography}
\end{document}